\documentclass[11pt,a4paper]{article}
\usepackage{amsfonts}
\usepackage{amsmath,amssymb}
\usepackage{epsfig,graphicx}

\input{tcilatex}

\begin{document}

\begin{center}
{\LARGE \ Emergent gauge theories and supersymmetry: a QED primer }

\bigskip \bigskip

\textbf{J.L.~Chkareuli}$^{1,2,3}$

$^{1}$\textit{Center for Elementary Particle Physics, Ilia State University,
0162 Tbilisi, Georgia\ \vspace{0pt}\\[0pt]
}

$^{2}$\textit{E. Andronikashvili} \textit{Institute of Physics, 0177
Tbilisi, Georgia\ }

$^{3}$\textit{Institute of High Energy Physics, Chinese Academy of Sciences,
Beijing 100049, China}

\textit{\bigskip }

\bigskip

\bigskip

\bigskip

\bigskip

\bigskip

\bigskip

\bigskip

\textbf{Abstract}

\bigskip
\end{center}

We argue that a generic trigger for photon and other gauge fields to emerge
as massless Nambu-Goldstone modes could be spontaneously broken
supersymmetry rather than physically manifested Lorentz violation. We
consider supersymmetric QED model extended by an arbitrary polynomial
potential of vector superfield that induces the spontaneous SUSY violation
in the visible sector. As a consequence, massless photon appears as a
companion of massless photino being Goldstone fermion state in tree
approximation. Remarkably, the photon masslessness appearing at tree level
is further protected against radiative corrections due to the simultaneously
generated special gauge invariance in the broken SUSY phase. Meanwhile,
photino being mixed with another goldstino appearing from a spontaneous SUSY
violation in the hidden sector largely turns into light pseudo-goldstino
whose physics seems to be of special interest.

\bigskip 

\thispagestyle{empty}\newpage

\section{Introduction and overview}

It is well known that spontaneous Lorentz invariance violation (SLIV) may
lead to an emergence of massless Nambu-Goldstone modes \cite{NJL} which are
identified with photons and other gauge fields appearing in the Standard
Model. This idea \cite{bjorken} supported by a close analogy with the
dynamical origin of massless particle excitations for spontaneously broken
internal symmetries has gained a new development \cite%
{cfn,jb,kraus,jen,bluhm} in recent years\footnote{%
Independently of the problem of the origin of local symmetries, Lorentz
violation in itself has attracted considerable attention as an interesting
phenomenological possibility which may be probed in direct Lorentz
non-invariant extensions of quantum electrodynamics and the Standard Model 
\cite{refs,2,3}.}.

In this connection, one important thing to notice is that, in contrast to
the spontaneous violation of internal symmetries, SLIV seems not to
necessarily imply a physical breakdown of Lorentz invariance. Rather, when
appearing in a gauge theory framework, this may eventually result in a
noncovariant gauge choice in an otherwise gauge invariant and Lorentz
invariant theory. In substance the SLIV ansatz, due to which the vector
field $A_{\mu }(x)$ develops a vacuum expectation value (vev) 
\begin{equation}
\left\langle A_{\mu }\right\rangle =n_{\mu }M  \label{vev1}
\end{equation}%
(where $n_{\mu }$ is a properly-oriented unit Lorentz vector, $n^{2}=n_{\mu
}n^{\mu }=\pm 1$, while $M$ is the proposed SLIV scale), may itself be
treated as a pure gauge transformation with a gauge function linear in
coordinates, $\omega (x)=$ $n_{\mu }x^{\mu }M$. From this viewpoint gauge
invariance in QED leads to the conversion of SLIV into gauge degrees of
freedom of the massless Goldstonic photon emerged. We will hereafter refer
to it as an "inactive" SLIV, as opposed to an "active" SLIV leading to
physical Lorentz violation (which may appear if gauge invariance is
explicitly broken, see below).

A good example for such a kind of the inactive SLIV is provided by the
nonlinearly realized Lorentz symmetry for underlying vector field $A_{\mu
}(x)$ through the length-fixing constraint%
\begin{equation}
A_{\mu }A^{\mu }=n^{2}M^{2}\text{ .}  \label{const}
\end{equation}%
This constraint in the gauge invariant QED framework was first studied by
Nambu a long ago \cite{nambu}, and in more detail in recent years \cite%
{az,kep,jej,urr,gra}. The constraint (\ref{const}) is in fact very similar
to the constraint appearing in the nonlinear $\sigma $-model for pions \cite%
{GL}, $\sigma ^{2}+\pi ^{2}=f_{\pi }^{2}$, where $f_{\pi }$ is the pion
decay constant. Rather than impose by postulate, the constraint (\ref{const}%
) may be implemented into the standard QED Lagrangian $L_{QED}$ through an
invariant Lagrange multiplier term%
\begin{equation}
L=L_{QED}-\frac{\lambda }{2}\left( A_{\mu }A^{\mu }-n^{2}M^{2}\right)
\label{lag}
\end{equation}%
provided that initial values for all fields (and their momenta) involved are
chosen so as to restrict the phase space to values with a vanishing
multiplier function $\lambda (x)$, $\lambda =0$ \footnote{%
Actually, due to an automatic conservation of the matter current in QED an
initial value $\lambda =0$ will remain for all time. In a general case, when
nonzero values of $\lambda $ are also allowed, it appears problematic to
have ghost-free theory theory with a positive Hamiltonian (for a detailed
discussion see \cite{vru}). It is also worth noting that, though the
Lagrange multiplier term formally breaks gauge invariance in the Lagrangian (%
\ref{lag}), this breaking, is in fact reduced to the nonlinear gauge choice (%
\ref{const}) in the restricted phase space mentioned above.}.

One way or another, the constraint (\ref{const}) means in essence that the
vector field $A_{\mu }$ develops the vev (\ref{vev1}) and Lorentz symmetry $%
SO(1,3)$ breaks down to $SO(3)$ or $SO(1,2)$ depending on whether the unit
vector $n_{\mu }$ is time-like ($n^{2}>0$) or space-like ($n^{2}<0$). The
point is, however, that, in sharp contrast to the nonlinear $\sigma $ model
for pions, the nonlinear QED theory, due to gauge invariance in the starting
Lagrangian $L_{QED}$, ensures that all the physical Lorentz violating
effects turn out to be non-observable. Actually, the nonlinear constraint (%
\ref{const}) implemented as a supplementary condition appears in essence as
a possible gauge choice for the vector field $A_{\mu }$, while the $S$%
-matrix remains unaltered under such a gauge convention. Indeed, this
nonlinear QED contains a plethora of Lorentz and $CPT$ violating couplings
when it is expressed in terms of the pure Goldstonic photon modes ($a_{\mu }$%
) according to the constraint condition (\ref{const})

\begin{equation}
A_{\mu }=a_{\mu }+n_{\mu }(M^{2}-n^{2}a^{2})^{\frac{1}{2}}\text{ , \ }n_{\mu
}a_{\mu }=0\text{ \ \ \ \ (}a^{2}\equiv a_{\mu }a^{\mu }\text{).}
\label{gol}
\end{equation}%
However, the contributions of all these Lorentz violating couplings to
physical processes completely cancel out among themselves. So, the inactive
SLIV inspired by the length-fixing constraint (\ref{const}) affects only the
gauge of the vector potential $A_{\mu }$ while leaving physical Lorentz
invariance intact, as was shown in tree \cite{nambu} and one-loop \cite{az}
approximations. Later a similar result was also confirmed for spontaneously
broken massive QED \cite{kep}, non-Abelian theories \cite{jej} and tensor
field gravity \cite{gra}.

From this point of view, emergent gauge theories induced by the inactive
SLIV mechanism are in fact indistinguishable from conventional gauge
theories. Their Goldstonic nature could only be seen when taking the gauge
condition of the length-fixing constraint type (\ref{const}). As to their
observational evidence, the only way for SLIV to cause physical Lorentz
violation would appear if gauge invariance in \ the theory considered were
really broken, rather than merely constrained by some gauge condition.

In this sense, a valuable alternative to the nonlinear QED model (\ref{lag})
might be a conventional QED type Lagrangian extended by an arbitrary vector
field potential energy terms which explicitly break gauge invariance. For a
minimal potential containing bilinear and quadrilinear vector field terms
one comes to the Lagrangian%
\begin{equation}
L=L_{QED}-\frac{\lambda }{4}\left( A_{\mu }A^{\mu }-n^{2}M^{2}\right) ^{2}
\label{lag2}
\end{equation}%
where $\lambda $ is now a coupling constant rather than the Lagrange
multiplier field. This potential being sometimes referred to as the
\textquotedblleft bumblebee\textquotedblright\ model (see \cite{bluhm} and
references therein) means in fact that the vector field $A_{\mu }$ develops
a constant background value (\ref{vev1}) causing again an appropriate
(time-like or space-like) Lorentz violation at a scale $M$. However, in
contrast to the nonlinear QED model with a directly imposed vector field
constraint (\ref{lag}), the bumblebee\ model contains an extra degree of
freedom which appears as a massive Higgs mode away from the potential
minimum. This allows SLIV to get active. Indeed, due to the presence of this
mode the model may lead to some physical Lorentz violation in terms of the
properly deformed dispersion relations for photon and matter fields involved
that appear from corresponding radiative corrections to their kinetic terms 
\cite{kraus}. However, in either case, whether SLIV is active or inactive,
it unavoidably leads to the generation of massless photons as vector
Nambu-Goldstone bosons.

Nevertheless, it may turn out that SLIV is not the only reason why massless
photons could dynamically appear, if spacetime symmetry is further enlarged.
In this connection, a special interest may be related to supersymmetry.
Actually, the situation is changed dramatically in the SUSY inspired models.
It appears that, while the nonlinear QED model with its inactive SLIV
successfully matches supersymmetry, the potential-extended QED models
(including the bumblebee\ model discussed above) leading to physical Lorentz
violation cannot be conceptually realized in the SUSY context\footnote{%
The point is that, in contrast to an ordinary vector field theory where all
kinds of terms with any power of the vector field squared, $(A_{\mu }A^{\mu
})^{n}$ ($n=1,2,...$), can be included into the Lagrangian in Lorentz
invariant way, SUSY theories only admit the bilinear term $A_{\mu }A^{\mu }$
in the vector field potential energy. This can readily be confirmed in SUSY
QED extended by an arbitrary polynomial potential of \ a general vector
superfield (Section 2). As a result, without a stabilizing high-linear (at
least quadrilinear, as in (\ref{lag2})) vector field terms, the
potential-based SLIV never occurs in SUSY theories.}. This allows to think
that physical Lorentz invariance is somewhat protected by SUSY and in this
sense a generic trigger for massless photons to dynamically appear could be
spontaneously broken supersymmetry itself rather than physically manifested
SLIV. To see how this idea might work we consider supersymmetric QED model
extended by an arbitrary polynomial potential of a general vector superfield
that induces the spontaneous SUSY violation. As a consequence, massless
photon emerges as a companion of massless photino being Goldstone fermion in
the broken SUSY phase in the visible sector (Section 2). Remarkably, this
masslessness appearing at tree level is further protected against radiative
corrections by the simultaneously generated special gauge invariance in the
Lagrangian at the SUSY\ breaking potential minimum (Section 3). Meanwhile,
photino being mixed with another goldstino appearing from a spontaneous SUSY
violation in the hidden sector largely turns into light pseudo-goldstino
whose physics seems to be of special interest. Some overall conclusion is
drawn in Section 4.

\section{Extended supersymmetric QED}

We now consider the supersymmetric QED extended by an arbitrary polynomial
potential of \ a general vector superfield $V(x,\theta ,\overline{\theta })$
whose a pure vector component $A_{\mu }$ is usually associated with a
photon. The corresponding Lagrangian can be written in the SUSY invariant
form as%
\begin{equation}
\mathcal{L}=L_{SQED}+\sum_{n=1}b_{n}V^{n}|_{D}  \label{slag}
\end{equation}%
where terms in this sum ($b_{n}$ are some constants) for a conventional
vector superfield parametrization\footnote{%
This parametrization is given by \cite{wess} $V(x,\theta ,\overline{\theta }%
)=C(x)+i\theta \chi (x)-i\overline{\theta }\overline{\chi }(x)+\frac{i}{2}%
\theta \theta S(x)-\frac{i}{2}\overline{\theta }\overline{\theta }S^{\ast
}(x)--\theta \sigma ^{\mu }\overline{\theta }A_{\mu }(x)+i\theta \theta 
\overline{\theta }\overline{\lambda ^{\prime }}(x)-i\overline{\theta }%
\overline{\theta }\theta \lambda ^{\prime }(x)+\frac{1}{2}\theta \theta 
\overline{\theta }\overline{\theta }D^{\prime }(x)$, where $S=M+iN$ , \ $%
\lambda ^{\prime }=\lambda +\frac{i}{2}\sigma ^{\mu }\partial _{\mu }%
\overline{\chi }$ \ and $D^{\prime }=D+\frac{1}{2}\partial ^{2}C$.} are
given by corresponding $D$-term expansions $V^{n}|_{D}$ into the component
fields. It can readily be checked that the first term in this expansion is
the known Fayet-Iliopoulos $D$-term, while other terms only contain
bilinear, trilinear and quadrilinear combination of the superfield
components $A_{\mu }$, $S$, $\lambda $ and $\chi $, respectively\footnote{%
Without loss of generality, we may restrict ourselves to the third degree
superfield polynomial in the Lagrangian $\mathcal{L}$ (\ref{slag}) to
eventually have a theory with dimesionless coupling constants for component
fields. However, for completeness sake, it seems better to proceed with a
general case.}. Actually, the higher-degree terms only appear for the scalar
field component $C(x)$. Expressing them all in terms of the $C$ field
polynomial%
\begin{equation}
P(C)=\sum_{n=1}\frac{n}{2}b_{n}C^{n-1}(x)  \label{pot}
\end{equation}%
and three of its derivatives with respect to the $C$ field 
\begin{equation}
P_{C}^{\prime }\equiv \frac{\partial P}{\partial C}\text{ , \ \ }%
P_{C}^{\prime \prime }\equiv \frac{\partial ^{2}P}{\partial C^{2}}\text{ , \
\ }P_{C}^{\prime \prime \prime }\equiv \frac{\partial ^{3}P}{\partial C^{3}}%
\text{ }  \label{dd}
\end{equation}%
one has for the whole Lagrangian $\mathcal{L}$ 
\begin{eqnarray}
\mathcal{L} &=&-\text{ }\frac{1}{4}F^{\mu \nu }F_{\mu \nu }+i\lambda \sigma
^{\mu }\partial _{\mu }\overline{\lambda }+\frac{1}{2}D^{2}  \notag \\
&&+\text{ }\left( D+\frac{1}{2}\partial ^{2}C\right) P+\left( \frac{1}{2}%
SS^{\ast }-\chi \lambda ^{\prime }-\overline{\chi }\overline{\lambda
^{\prime }}-\frac{1}{2}A_{\mu }A^{\mu }\right) P_{C}^{\prime }  \notag \\
&&+\text{ }\frac{1}{2}\left( \frac{i}{2}\overline{\chi }\overline{\chi }S-%
\frac{i}{2}\chi \chi S^{\ast }-\chi \sigma ^{\mu }\overline{\chi }A_{\mu
}\right) P_{C}^{\prime \prime }+\frac{1}{8}(\chi \chi \overline{\chi }%
\overline{\chi })P_{C}^{\prime \prime \prime }  \label{lag3}
\end{eqnarray}%
where we properly calculated $D$-terms for the corresponding powers of the
superfield $V(x,\theta ,\overline{\theta })$ in the polynomial in (\ref{slag}%
) and collected like field couplings. Also, for more clarity, we still
omitted matter superfields in the model (see some discussion below). As one
can see, extra degrees of freedom related to the $C$ and $\chi $ component
fields in a general vector superfield $V(x,\theta ,\overline{\theta })$
appear through the potential terms in (\ref{lag3}) rather than from the
properly constructed supersymmetric field strengths, as is only appeared for
the vector field $A_{\mu }$ and its gaugino companion $\lambda $.

Note that all terms in the sum in (\ref{slag}) except Fayet-Iliopoulos $D$%
-term\ explicitly break gauge invariance. However, as we will see in the
next section, the special gauge invariance constrained by some gauge
condition will be recovered in the Lagrangian in the broken SUSY phase.
Furthermore, as is seen from (\ref{lag3}), the vector field $A_{\mu }$ may
only appear with bilinear mass term in the polynomially extended superfield
Lagrangian (\ref{slag}). This means that, in contrast to the non-SUSY theory
case (where, apart from the vector field mass term, some high-linear
stabilizing\ terms necessarily appear in similar polynomially extended
Lagrangian), Lorentz invariance is still preserved. Actually, only
supersymmetry appears to be spontaneously broken, as mentioned above.

Indeed, varying the Lagrangian $\mathcal{L}$ with respect to the $D$ field
we come to 
\begin{equation}
D=-P(C)  \label{d}
\end{equation}%
that finally gives the following potential energy for the field system
considered 
\begin{equation}
U(C)=\frac{1}{2}[P(C)]^{2}\text{ .}  \label{pot1}
\end{equation}%
being solely determined by the polynomial of the scalar field component $%
C(x) $ of the superfield $V(x,\theta ,\overline{\theta })$. The potential (%
\ref{pot1}) may lead to the spontaneous SUSY breaking provided that the
polynomial $P$ (\ref{pot}) has no real roots, while its first derivative
has, 
\begin{equation}
P\neq 0\text{ , \ }P_{C}^{\prime }=0.\text{\ }  \label{der}
\end{equation}%
This requires $P(C)$ to be an even degree polynomial with properly chosen
coefficients $b_{n}$ in (\ref{pot}) that will force its derivative $%
P_{C}^{\prime }$ to have at least one root, $C=C_{0}$, in which the
potential (\ref{pot1}) is minimized. Therefore, supersymmetry is
spontaneously broken and the $C$ field acquires the vev 
\begin{equation}
\left\langle C\right\rangle =C_{0}\text{ , \ }P_{C}^{\prime }(C_{0})=0\text{
.}  \label{vvv}
\end{equation}%
As an immediate consequence, that one can readily see from the Lagrangian $%
\mathcal{L}$ (\ref{lag3}), a massless photino $\lambda $ being Goldstone
fermion in the SUSY broken phase make all the other component fields in the
superfield $V(x,\theta ,\overline{\theta }),$ including the photon, to also
become massless. However, the question then arises whether this masslessness
of photon will be stable against radiative corrections\footnote{%
These corrections include as ordinary corrections appearing in a
conventional supersymmetric QED, so corrections related to new degrees of
freedom emerging in a general vector superfield $V(x,\theta ,\overline{%
\theta })$ in terms of the revived component fields $C$ and $\chi $. All
these corrections appear relativistically invariant since, as was mentioned
above, physical Lorentz invariance is preserved in the basic Lagrangian $%
\mathcal{L}$ (\ref{lag3}).}, since gauge invariance is explicitly broken in
the Lagrangian (\ref{lag3}). We show in the next section that it could be
the case if the vector superfield $V(x,\theta ,\overline{\theta })$ would
appear to be properly constrained.

Before proceeding further, note that we have not yet included matter
superfields in the model. In their presence, the spontaneous SUSY breaking
in the visible sector we have used above should be properly combined with a
spontaneous SUSY violation in the hidden sector to evade the supertrace sum
rule \cite{wess} for masses of basic fermions and their superpartners.
Actually, this sum rule is acceptably relaxed when taking into account large
radiative corrections to masses of supersymmetric particles, as typically
appears in gauge-mediated SUSY breaking models \cite{wess}. This changes the
simplified picture discussed above: the strictly massless fermion eigenstate
appears to be some mix of the visible sector photino $\lambda $ and the
hidden sector\ goldstino rather than the pure photino state. In the
supergravity context, one linear combination of them is eaten through the
super-Higgs mechanism to form the longitudinal component of the gravitino,
while their orthogonal combination, which may be referred to as a
pseudo-goldstino, gets some mass\footnote{%
The possibility that the Standard Model visible sector might also break
supersymmetry thus giving rise to similar pseudo-goldstino states was also
considered, though in a different context, in \cite{vis,tha}.}. One may
generally expect that SUSY is much stronger broken in the hidden sector than
in the visible one that means the pseudo-goldstino is largely given by the
pure photino state $\lambda $. These states seem to be of special
observational interest in the model that, apart from some indication of the
QED emergence nature, may shed a light on \ the SUSY breaking physics.

\section{Constrained vector superfield}

We have seen above that the vector field $A_{\mu }$ may only appear with
bilinear mass terms in the polynomially extended Lagrangian (\ref{lag3}).
Hence it follows that potential-based \ models, particularly the bumblebee
model mentioned above (\ref{lag2}) with nontrivial vector field potential
containing both a bilinear mass term and a quadrilinear stabilizing term,
can in no way be realized in the SUSY context. Meanwhile, the nonlinear QED
model, as will become clear below, successfully matches supersymmetry.

Let us notice first that instead of gauge symmetry broken in the extended
QED Lagrangian (\ref{lag3}) some special gauge invariance is recovered in (%
\ref{lag3}) at the SUSY\ breaking minimum of the potential (\ref{pot1}). We
will expand the action around the vacuum (\ref{vvv}) by writing 
\begin{equation}
C(x)=C_{0}+c(x)  \label{cc}
\end{equation}%
that gives for the $C$ field polynomial $P(C)$ (\ref{pot}) and its
derivatives (\ref{dd}) to the lowest order in the Higgs-like field $c(x)$ 
\begin{eqnarray}
P(C) &\simeq &P(C_{0})+\frac{1}{2}P_{C}^{\prime \prime }(C_{0})c^{2}\text{ ,
\ }P_{C}^{\prime }(C)\simeq P_{C}^{\prime \prime }(C_{0})c\text{ , \ } 
\notag \\
P_{C}^{\prime \prime }(C) &\simeq &P_{C}^{\prime \prime
}(C_{0})+P_{C}^{\prime \prime \prime }(C_{0})c\text{ , \ }P_{C}^{\prime
\prime \prime }(C)\simeq P_{C}^{\prime \prime \prime }(C_{0})+P_{C}^{\prime
\prime \prime \prime }(C_{0})c\text{\ }  \label{app}
\end{eqnarray}%
with $P_{C}^{\prime }(C_{0})=0$ taken at the minimum point, as is determined
in (\ref{der}). Now, combining the equations of motion for $c(x)$ and for
some other component field, say $S(x)$, both derived by varying the
Lagrangian (\ref{lag3}), one has 
\begin{equation}
\frac{1}{2}SS^{\ast }-\chi \lambda ^{\prime }-\overline{\chi }\overline{%
\lambda ^{\prime }}-\frac{1}{2}A_{\mu }A^{\mu }=O(c,c\partial ^{2}c)\text{ ,
\ \ }\chi \chi =O(c)  \label{sa}
\end{equation}%
where we have used approximate equalities (\ref{app}) with typically
sizeable values of all $P(C_{0})$, $P_{C}^{\prime \prime }(C_{0})$, $%
P_{C}^{\prime \prime \prime }(C_{0})$, $P_{C}^{\prime \prime \prime \prime
}(C_{0})$ taken at the minimum point $C_{0}$. So, at the SUSY\ breaking
minimum ($c\rightarrow 0$) we come to the constraints which are put on the $%
V $ superfield components 
\begin{equation}
C=C_{0},\text{ }\chi =0,\text{\ \ \ }A_{\mu }A^{\mu }=|S|^{2}\text{\ }
\label{const1}
\end{equation}%
that also determine the corresponding $D$-term (\ref{d}), $D=-P(C_{0}),$ for
the spontaneously broken supersymmetry.

Another, more exact, way to keep the whole theory at the SUSY\ breaking
minimum, thus eliminating Higgs-like mode $c(x)$ forever, is to properly
constrain the vector superfield $V(x,\theta ,\overline{\theta })$ from the
outset. This can be done, by analogy with constrained vector field in the
nonlinear QED model (\ref{lag}), through the following SUSY\ invariant
Lagrange multiplier term 
\begin{equation}
\mathcal{L}^{\prime }=\mathcal{L}+\frac{1}{2}\Lambda (V-C_{0})^{2}|_{D}
\label{ext}
\end{equation}%
where $\Lambda (x,\theta ,\overline{\theta })$ is some auxiliary vector
superfield, while $C_{0}$ is the constant background value of the $C$ field
for which potential $U(C)$ has the SUSY breaking minimum (\ref{vvv}).

We further find for the Lagrange multiplier term in (\ref{ext}) that
(denoting $\widetilde{C}\equiv C-C_{0}$)%
\begin{eqnarray}
\Lambda V^{2}|_{D} &=&{\large C}_{\Lambda }\left[ \widetilde{C}D^{\prime
}+\left( \frac{1}{2}SS^{\ast }-\chi \lambda ^{\prime }-\overline{\chi }%
\overline{\lambda ^{\prime }}-\frac{1}{2}A_{\mu }A^{\mu }\right) \right]  
\notag \\
&&+\text{ }{\large \chi }_{\Lambda }\left[ 2\widetilde{C}\lambda ^{\prime
}+i(\chi S^{\ast }+i\sigma ^{\mu }\overline{\chi }A_{\mu })\right] +%
\overline{{\large \chi }_{\Lambda }}[2\widetilde{C}\overline{\lambda
^{\prime }}-i(\overline{\chi }S-i\chi \sigma ^{\mu }A_{\mu })]  \notag \\
&&+\text{ }\frac{1}{2}{\large S}_{\Lambda }\left( \widetilde{C}S^{\ast }+%
\frac{i}{2}\overline{\chi }\overline{\chi }\right) +\frac{1}{2}S_{\Lambda
}^{\ast }\left( \widetilde{C}S-\frac{i}{2}\chi \chi \right)   \notag \\
&&+\text{ }2{\large A}_{\Lambda }^{\mu }(\widetilde{C}A_{\mu }-\chi \sigma
_{\mu }\overline{\chi })+2{\large \lambda }_{\Lambda }^{\prime }(\widetilde{C%
}\chi )+2\overline{{\large \lambda }_{\Lambda }^{\prime }}(\widetilde{C}%
\overline{\chi })+\frac{1}{2}{\large D}_{\Lambda }^{\prime }\widetilde{C}^{2}
\label{lm1}
\end{eqnarray}%
where 
\begin{equation}
{\large C}_{\Lambda },\text{ }{\large \chi }_{\Lambda },\text{ }{\large S}%
_{\Lambda },\text{ }{\large A}_{\Lambda }^{\mu },\text{ }{\large \lambda }%
_{\Lambda }^{\prime }={\large \lambda }_{\Lambda }+\frac{i}{2}\sigma ^{\mu
}\partial _{\mu }\overline{{\large \chi }_{\Lambda }},\text{ }{\large D}%
_{\Lambda }^{\prime }={\large D}_{\Lambda }+\frac{1}{2}\partial ^{2}{\large C%
}_{\Lambda }  \label{comp}
\end{equation}%
are the standard$^{4}$ component fields of the Lagrange multiplier
superfield $\Lambda (x,\theta ,\overline{\theta })$. Now varying the whole
Lagrangian (\ref{ext}) with respect to these nondynamical fields and
properly combining their equations of motion 
\begin{equation}
\frac{\partial \mathcal{L}^{\prime }}{\partial \left( {\large C}_{\Lambda },%
{\large \chi }_{\Lambda },{\large S}_{\Lambda },{\large A}_{\Lambda }^{\mu },%
{\large \lambda }_{\Lambda },{\large D}_{\Lambda }\right) }=0
\end{equation}%
we come again to the constraints (\ref{const1}) imposed on the vector
superfield components. As before in non-SUSY case (\ref{lag}), to have
ghost-free theory with a positive Hamiltonian the initial values for all
component fields of superfield $V(x,\theta ,\overline{\theta })$ are chosen
so as to restrict phase space to values with the vanishing component fields (%
\ref{comp}) of the multiplier superfield $\Lambda (x,\theta ,\overline{%
\theta })$.

Remarkably, one way or another, we have come for the constrained vector
superfield $V(x,\theta ,\overline{\theta })$ to almost the same physical
states, photino and photon, as in supergauge multiplet of conventional
supersymmetric QED \cite{wess}. Actually, photino field $\lambda $ appears
to be constraint-free, while the vector potential $A_{\mu }$ is only
constrained by the condition relating it to the nondynamical $S$ field. One
can now readily confirm that, as a consequence of the spontaneous SUSY
violation inducing the constraints (\ref{const1}), some special gauge
invariance is in fact recovered in the Lagrangian (\ref{slag}). First of
all, this violation provides the tree level photon masslessness,\ as is
clearly seen in the Lagrangian (\ref{lag3}) when the potential minimum
condition $P_{C}^{\prime }(C_{0})=0$ is applied. The rest of gauge
noninvariance caused by a general superfield polynomial in (\ref{slag}) is
simply reduced to the nonlinear gauge choice $A_{\mu }A^{\mu }=|S|^{2}$ in a
virtually gauge invariant theory since extra degrees of freedom in terms of
the $C$ and $\chi $ component fields are also eliminated. Taking the $S$
field to be some constant background field we come to the SLIV\ constraint (%
\ref{const}) being in an ordinary nonlinear QED discussed above. As is seen
from this gauge, one may only have a time-like SLIV in the SUSY framework
but never a space-like one. There also may be a light-like SLIV, if the $S$
field vanishes\footnote{%
Indeed, this case, first mentioned in \cite{nambu}, may also mean
spontaneous Lorentz violation with a nonzero vev $<A_{\mu }>$ $=(\widetilde{M%
},0,0,\widetilde{M})$ and Goldstone modes $A_{1,2}$ and $(A_{0}+A_{3})/2$\ $-%
\widetilde{M}.$ The "effective" Higgs mode $(A_{0}-A_{3})/2$ can be then
expressed through Goldstone modes so that the light-like condition $A_{\mu
}^{2}=0$ is satisfied.}. All of them are in fact inactive SLIV models in
which physical Lorentz invariance is left intact. So, any possible choice
for the $S$ field only leads to the particular gauge choice for the vector
field $A_{\mu }$ in an otherwise gauge invariant theory. Thus, the massless
photon appearing first as a companion of massless photino (being Goldstone
fermion in the visible broken SUSY phase) remains massless due to this
recovering gauge invariance in the emergent SUSY QED. At the same time, the
"built-in" nonlinear gauge condition (\ref{const1}) gives rise to treat the
photon as vector Goldstone boson induced by the inactive SLIV.

\section{Conclusion}

We have attempted to find some supersymmetric analogue of the emergent QED
inspired by spontaneous Lorentz invariance violation (SLIV). As discussed
above, SLIV\ in a vector field theory framework may be inactive as in the
nonlinear QED model (\ref{lag}), or active as in potential-extended QED
models (\ref{lag2}) leading to physical Lorentz violation. However, in
either case SLIV unavoidably leads to the generation of massless
Nambu-Goldstone modes which are identified with photons and other gauge
fields appearing in the Standard Model. Nevertheless, it may turn out that
SLIV is not the only reason why massless photons could dynamically appear,
if spacetime symmetry is further enlarged. In this connection, a special
interest is related to supersymmetry that we argued in this note by the
example of supersymmetric QED\footnote{%
It is worth noting that all the basic arguments related to the present QED
example can be extended to the Standard Model.}. Actually, while there are a
few papers in the literature on Lorentz noninvariant extensions of
supersymmetric models (for some interesting ideas, see \cite{lor} and
references therein), an emergent gauge theory in a SUSY context has been
considered for the first time.

In substance, the situation is changed dramatically in SUSY models: between
two basic SLIV\ versions mentioned above, SUSY inevitably chooses the
inactive SLIV case. Indeed, while the nonlinear QED model with its inactive
SLIV successfully matches supersymmetry, the potential-extended QED models
with an actual Lorentz violation can never occur in the SUSY context. This
allows to think that a generic trigger for massless photons to appear could
be spontaneously broken supersymmetry itself rather than physically
manifested spontaneous Lorentz violation. For an explicit demonstration we
considered supersymmetric QED model extended by an arbitrary polynomial
potential of massive vector superfield that induces the spontaneous SUSY
violation. As a consequence, massless photon emerges as a companion of
massless photino being Goldstone fermion in the broken SUSY phase.

The photon masslessness appearing at tree level is further protected against
radiative corrections by the simultaneously generated special gauge
invariance. This invariance is only restricted by the nonlinear gauge
condition put on vector field values, $A_{\mu }A^{\mu }=|S|^{2}$ with the
nondynamical $S$ field chosen as some arbitrary constant background field
(including the vanishing one) in the theory. The point, however, is that
this nonlinear gauge condition happens at the same time to be the SLIV type
constraint which treats in turn the physical photon as the Lorentzian
Goldstone mode. So, figuratively speaking, the photon passes through three
evolution stages being initially the massive vector field component of a
general vector superfield (\ref{lag3}), then the three-level massless
companion of the Goldstonic photino $\lambda $ in the broken SUSY stage (\ref%
{vvv}) and finally the generically massless state as the emergent Lorentzian
mode in the inactive SLIV stage (\ref{const1}).

Meanwhile, photino being mixed with another goldstino appearing from a
spontaneous SUSY violation in the hidden sector largely turns into light
pseudo-goldstino whose physics seems to be of special interest. As argued in 
\cite{tha}, if the SUSY\ visible sector possesses $R$-symmetry then the
pseudo-goldstino mass is protected up to ($R$-violating) supergravity
effects, and eventually the same region of parameter space simultaneously
may solve both gravitino and pseudo-goldstino overproduction problems in the
early universe. Apart from cosmological implications, many other sides of
new physics related to pseudo-goldstinos appearing through the multiple SUSY
breaking were also studied recently (see \cite{vis,tha,gol} and references
therein). The point is, however, that there have been exclusively used
non-vanishing $F$-terms as the only mechanism of visible SUSY breaking in
models considered. In this connection, our pseudo-goldstonic photinos caused
by non-vanishing $D$-terms in the visible SUSY sector may lead to somewhat
different observational consequences. We are going to return to this
interesting issue elsewhere.

\section{Acknowledgments}

I appreciate the kind hospitality shown during my visit to Institute of High
Energy Physics, Chinese Academy of Sciences where part of this work was
carried out. Discussions with Colin Froggatt, Oleg Kancheli, Archil
Kobakhidze, Yi Liao, Rabi Mohapatra, Holger Nielsen and Zhi-Zhong Xing, as
well as with participants of the SUSY 2012 conference (13-18 August 2012,
Beijing, China) where some preliminary version of this work was reported are
also gratefully acknowledged.

\end{document}